\documentclass[apj]{emulateapj}

\usepackage{graphicx}
\usepackage{dcolumn}
\usepackage{bm}
\usepackage{rotating}

\usepackage[]{natbib}
\bibpunct{(}{)}{,}{a}{}{,}
\begin{document}


\title{Inferring physical parameters of compact stars from their f-mode gravitational wave signals}

\author{H. K. Lau\footnote{Present address: Department of Physics,
University of Toronto, 60 St. George St., Toronto, M5S 1A7,
Ontario, Canada; email: kero.lau@utoronto.ca},
 P. T. Leung\footnote{email: ptleung@phy.cuhk.edu.hk}~ and L. M. Lin\footnote{email: lmlin@phy.cuhk.edu.hk}}
 \affil{Department of Physics and Institute of Theoretical Physics,\\
 The Chinese University of Hong Kong, Shatin, Hong Kong SAR, China.}

\date{\today}


\begin{abstract}
We propose here a robust scheme to infer the physical parameters
of compact stars from their f-mode gravitational wave signals. We
first show that the frequency and the damping rate of f-mode
oscillation of compact stars can be expressed in terms of
universal functions of stellar mass and moment of inertia. By
employing the universality in the f-mode one can then infer
accurate values of the mass, the moment of inertia and the radius
of a compact star. In addition, we demonstrate that our new scheme
works well for both realistic neutron stars and quark stars, and
hence provides a unifying way to infer the physical parameters of
compact stars.

\end{abstract}
\def\tomega{ \tilde{\omega} }
\def\tOmega{ \tilde{\Omega} }
\def\tr{ \tilde{r} }
\def\tx{ \tilde{r}_* }
\def\tV{ \tilde{V} }
\def\tpsi{ \tilde{\psi} }
\def\trho{ \tilde{\rho} }
\def\tP{ \tilde{P} }
\def\tR{ \tilde{R} }
\def\tX{ \tilde{R}_* }
\def\tm{ \tilde{m} }
\def\tphi{ \tilde{\nu} }
\def\tlam{ \tilde{\lambda}}
\def\tepsilon { \tilde{\epsilon}}
\def\tHO{ \tilde{H}_0}
\def\tHI{ \tilde{H}_1}
\def\tK{ \tilde{K}}
\def\tW{ \tilde{W}}
\def\tX{ \tilde{X}}
\def\tV{ \tilde{V}}
\def\tgamma{ \tilde{\gamma}}
\def\ta{ \tilde{a}}
\def\tb{ \tilde{b}}
\def\tg{ \tilde{g}}
\def\th{ \tilde{h}}
\def\tk{ \tilde{k}}
\def\a{ {\rm a}}
\def\c{ {\rm c}}
\def\e{ {\rm e}}
\def\p{ {\rm p}}
\def\f{ {\rm f}}
\def\d{ {\rm d}}
\def\i{ {\rm i}}
\def\r{ {\rm r}}

\keywords{gravitational waves --- stars: neutron ---  stars:
oscillations (including pulsations) --- equation of state ---
relativity
 }

\maketitle

\section{\label{intro}Introduction}
Ever since the first prediction of neutron stars (NS) by
\citet{Baade1934a}, NS have become a major test bed of various
theories of dense matter and nuclear matter. The extreme high
pressure and density achievable inside NS offers an ideal and
unique environment to infer the equation of state (EOS) of nuclear
matter from astronomical observations. On the other hand, the
possible existence of quark stars (QS) would provide a strong
support to a new form of matter, namely, the quark matter
\citep{Itoh70, Bodmer71, Witten}. Moreover, the size and the mass
of QS could also cast light on the properties of quark matter.
Thus, extracting useful information and constraints about nuclear
and particle physics from astronomical observations of compact
stars like NS and QS has become  an active direction in
astrophysics \citep[see, e.g.,][for reviews]
{lattimer2007nso,haensel2007nse}.

Since the density in the core of a NS (or QS) can be several times
the normal nuclear density, the EOS there is rather uncertain.
Various physical phenomena including pion condensation, hyperon
formation and deconfined quark matter have been considered by
theoretical nuclear physicists and sophisticated techniques of
many-body calculations have been developed to yield a bunch of EOS
intended for the description of the deep interior of compact
stars. It is standard practice for astrophysicists to construct
compact stars with these EOS for nuclear matter and establish
various relationships among different  physical quantities which
could be obtained from astronomical observations.
 Motivated by
the possibilities of constraining EOS of nuclear matter and
extracting physical parameters of compact stars as well,
researchers in astrophysics and nuclear physics have been actively
seeking for EOS-dependent and EOS-independent relationships  by
examining the physical characteristics of compact stars
constructed with different EOS \citep[see,
e.g.,][]{Lattimer:2001,Bejger:2002p8392,Lattimer:2005p7082,
bejger2005cdm,lattimer2007nso,haensel2007nse}. For example,
\citet{Lattimer:2001} compared the structure of compact stars of
various kinds and discovered several empirical relationships
connecting different physical characteristics of a star, such as
the mass, the radius, the moment of inertia, and the mass
distribution function as well. Such relationships can be applied
to infer the physical attributes of a compact star, including its
EOS, from astronomical observations.

The main objective of the present paper is to unveil the
universality embedded in the pulsation frequencies of
non-rotating compact stars and to exploit these findings to infer
the physical characteristics of NS or QS (e.g., mass, radius,
moment of inertia and EOS) from their oscillation spectra. The
pulsations of compact stars are an interesting and popular topic
in its own right, for gravitational waves (GW) can be generated
in such processes \citep[see][for a seminal exposition of the
topic]{Thorne}. It is indisputable that the detection of GW would
be a major milestone in general relativity and astrophysics.
Several Earth-based GW interferometric detectors such as LIGO,
VIRGO, GEO600 and TAMA300 have been operating. While the current
detectors are still not sensitive enough to detect GW directly,
interesting upper limits have been placed on either the GW
strains or event rate for several potential astrophysical sources
\citep[see, e.g., ][for the results obtained from the latest
science runs of LIGO]{LIGO_S4_BHringdown,LIGO_S5_LowMass,
LIGO_S5_pulsars}.

Owing to the energy carried away by GW, pulsations of compact
stars are damped harmonic oscillations, which are analyzed in
terms of quasi-normal modes (QNM) \citep[see,
e.g.][]{Press_1971,Leaver_1986,rmp,Kokkotas_rev}. Each QNM is
characterized by a complex eigenfrequency $\omega=\omega_\r+
i\omega_\i$ and has a time dependence $\exp(i\omega t)$,
displaying exponential decay with a damping time $\tau\equiv
1/\omega_\i$. Several attempts have been made to relate
$\omega_\r$ and $\omega_\i$ (or $\tau$) to the mass $M$ and the
radius $R$ of a compact star and some universal relationships
which are, to certain degree of accuracy, EOS-independent have
been found
\citep{Andersson1996,Andersson1998,Ferrari,Ferrari_prd,Tsui05:1}.
However, most of  these relationships fail to describe QNM of QS,
which are stiff and self-bound.

On the other hand, \citet{Lattimer:2005p7082} found an empirical
relation approximately expressing the moment of inertia in terms
of the mass and the radius. Such a relationship is universal as
long as the EOS under consideration is not too soft (e.g., hyperon
matter) or too stiff (e.g., quark matter). They proposed to use
the universality to estimate the radius of a compact star from its
mass and moment of inertia, with the latter two quantities being
measurable for double pulsar systems by considering the spin-orbit
coupling effect in general relativity \citep{Lattimer:2005p7082}.

Motivated by the discovery of \citet{Lattimer:2005p7082}, we
propose here to investigate the correspondence between the QNM
frequency (both $\omega_\r$ and $\omega_\i$) of the (fundamental)
f-mode oscillations, $M$ and the moment of inertia $I$ of compact
stars. We show that there exist EOS-independent universal
relationships between them, namely equations (\ref{lau_r}) and
(\ref{lau_i}), which are accurate up to a few percent or better
and work nicely for QS as well. Based on equations (\ref{lau_r})
and (\ref{lau_i}), a feasible scheme is in turn established to
determine the mass and the moment of inertia of a compact star
once its f-mode frequency is found from GW observations. As the
radius $R$ of a NS (or QS) is approximately expressible in terms
of the mass and the moment of inertia, our scheme can also yield
good estimate of the radius and hence imposing constraints on the
EOS \citep{Bejger:2002p8392,Lattimer:2005p7082}.

The outline of the paper is as follows. Section 2 is a brief
review on the the behavior of f-mode oscillations of compact stars
\citep{Tsui05:1} and the universal relationships among the mass,
the radius and the moment of inertia of compact stars
\citep{Bejger:2002p8392,Lattimer:2005p7082}.  In Section 3  we
introduce an accurate universal relation between the scaled
frequency $M \omega$ and the physical quantity $\sqrt{M^3/I}$.  In
Section 4 we establish a feasible and robust scheme to apply our
findings reported in Section 3 to invert the mass, the moment of
inertia, and in turn the radius of a compact star if its f-mode GW
signal is detected. We finally summarize our paper in Section 5
with a brief discussion. Unless otherwise noted, we adopt
geometric units with $G=c=1$ throughout the whole paper, and use
kilometers as the unit of lengths.

\section{GW and moment of inertia of compact stars}
As mentioned in Section 1, the oscillation modes of compact stars
are damped due to the emission of GW. Hence, they are called
quasi-normal modes, which can be further divided into polar and
axial classes in each angular momentum sector according to the
property of relevant metric and fluid perturbations under
inversion. For typical non-rotating compact stars, the polar
class includes (fundamental) f-mode, (gravity) g-mode, and
(pressure) p-mode and (spacetime) w-mode, while for  non-rotating
perfect fluid stars the axial class contains only the w-modes
\citep[see, e.g.,][for a review on the property of these
modes]{Kokkotas_rev}. The methods for evaluation of QNM of compact
stars are well documented in several papers and reviews
\citep[see,
e.g.,][]{Lindblom_1983,Lindblom-1985,Chandrasekhar1,Kokkotas_rev}.

The study of QNM of compact stars has a long history dating back
to the pioneering work of \citet{Thorne}. It is by now well
established that the study of GW mode spectra of compact stars
could provide useful information about the internal structure of
the stars \citep[see,
e.g.,][]{Andersson1996,Andersson1998,Kokkotas_2001,Ferrari,Ferrari_prd,Tsui05:3,Tsui05:1,QS_1,QS_2,Wong2009}.
As it is expected that the direct detection of GW will be realized
in the near future, different proposals have been put forward to
determine the physical parameters (e.g., mass, radius and EOS) of
a compact star from its GW spectra. In the present paper we focus
our attention on an EOS-independent universal behavior of the
quadrupolar (angular momentum index $l=2$) f-mode and show that
such universality can lead to accurate determination of the mass
and the moment of inertia of a compact star (NS or QS). Combining
our results with equations (\ref{Haensel_uni}), (\ref{Lat}) and
(\ref{Haensel_QS}), one can also find an approximate value of the
radius.

The reason why we place special emphasis on the f-mode is that the
relatively low frequency ($\sim 1-3$ kHz) of the f-mode makes it
more promising to be detectable by GW detectors than other modes.
For example, considering a compact star 10 kpc from us,
\citet{Kokkotas_2001} estimated that the energy required in the
f-mode in order to lead to a detection with signal-to-noise ratio
of 10 by the LIGO detector is $4.9\times 10^{-5} M_{\odot} c^2$.
For comparison, in order to achieve the same signal-to-noise
ratio, the energy in the p-mode (w-mode) must be $4.0 \times
10^{-3} M_{\odot} c^2$ ($6.8 \times 10^{-2} M_{\odot} c^2$). The
energy requirement is much relaxed for the advanced LIGO detector
and the same signal-to-noise ratio is achieved if the energy in
the f-mode is $8.7\times 10^{-7} M_{\odot} c^2$. Since the total
energy released in violent astronomical events such as supernova
explosions could be as much as $10^{-2}\, M_{\odot} c^2$ in some
optimistic estimates \citep{Kokkotas_2001}, it seems plausible
that the f-mode QNM signal of a fiercely pulsating compact star
inside our galaxy could be detected by future GW detectors,
especially when the narrow-banding capability is available to
enhance the sensitivity of the detectors in the high-frequency
region at around 1 kHz.

There have been several attempts to search for an approximately
EOS-independent behavior of the f-mode
\citep{Andersson1996,Andersson1998,Ferrari,Ferrari_prd,Tsui05:1}.
In particular, \citet{Tsui05:1} showed that the scaled frequency
$M \omega$ for the f-mode of realistic NS is essentially a
universal function of the compactness $M/R$, which can be
approximated by the quadratic formulae:
\begin{eqnarray}\label{leung_uni}
M\omega_\r &=& 0.15 \left(\frac{M}{R}\right)^2 + 0.56 \left(\frac{M}{R}\right) - 0.02 \,, \label{leung1}\\
\frac{M\omega_\i}{10^{-4}}  &=& -5.8 \left(\frac{M}{R}\right)^2 +
6.7 \left(\frac{M}{R}\right) - 0.62   \,. \label{leung2}
\end{eqnarray}
These two formulae are good approximation for most NS. However,
the f-modes of QS  do not follow equations (\ref{leung1}) and
(\ref{leung2}). Instead, they are approximated better by another
quadratic fit. The difference between the behaviors of the f-modes
of NS and QS are expected because the density profile of a QS is
nearly uniform, which is very different from that of a NS.

On the other hand,  \citet{Bejger:2002p8392} and
\citet{Lattimer:2005p7082} discovered  some empirical
relationships relating $I$ and compactness $M/R$. For NS,
\citet{Bejger:2002p8392} found that
\begin{eqnarray}\label{Haensel_uni}
\tilde{I} = \Bigg\{\begin{array}{ll} x/(0.1+2x) ~~& \textrm{if $x\le 0.1$,}\\
{2}(1+5 x)/{9} & \textrm{if $x>0.1$,}
\end{array}
\end{eqnarray}
where the scaled moment of inertia $\tilde{I} = I/MR^2$ and $x=
(M/M_{\odot})({\rm km}/R)$. \citet{Lattimer:2005p7082} also
proposed an alternative expression:
\begin{equation}\label{Lat}
\tilde{I} = 0.237 (1+4.2x+90x^4)~,
\end{equation}
which yields better agreement  for NS with higher compactness.
However, both equations (\ref{Haensel_uni}) and (\ref{Lat}) fail
to describe the case of QS. Instead, \citet{Bejger:2002p8392}
found another relationship for QS:
\begin{equation}\label{Haensel_QS}
\tilde{I}=\frac{2}{5}(1+x)~.
\end{equation}
 The distinction between the behaviors of NS
and QS is also expected and attributable to the difference in the
stiffness between nuclear and quark matters.

Assuming that both $I$ and $M$ are obtained from observations,
\citet{Lattimer:2005p7082} further argued that the radius $R$ can
be found by solving these equation. However, two remarks about
this method are in order. First, for the NS case the data points
corresponding to stars characterized by different EOS are rather
scattered and may deviate appreciably from equations
(\ref{Haensel_uni}) and (\ref{Lat}) . As a consequence, the value
of $R$ obtained in this way is subjected to uncertainties
depending on the EOS. Second, additional information has to be
gathered to distinguish between the data of QS and NS beforehand.
Regardless of these remarks, it is observed that for a fixed EOS
the scaled moment of inertia increases monotonically with the
compactness, indicating a one-one correspondence between these two
quantities. In the following discussion we shall consider the
feasibility to replace the compactness used in equations
(\ref{leung1}) and (\ref{leung2}) with the moment of inertia.


\section{Universality in f-mode oscillations  \label{moi uni}}
It is interesting to note from the reviews sketched in Sections 2
that both the scaled moment of inertia $I/MR^2$ and scaled f-mode
frequency $M \omega$ are rather insensitive to EOS and are, within
certain accuracy, universal functions of compactness
\citep{Bejger:2002p8392,Lattimer:2005p7082,Tsui05:1}. Such
coincidence is interesting, reflecting the fact that compactness
is a useful parameter characterizing a compact star. However, it
is also noticed that the scaled f-mode frequencies (or the scaled
moments of inertia)  of NS and QS follow two different universal
behaviors. The disparity in the universal behaviors of NS and QS
is attributable to the difference in their density profiles.

In the present paper we propose to examine the possibility of
using the moment of inertia as an independent parameter, instead
of the compactness, to characterize the f-mode spectra and in turn
to search for any universality underlying these spectra. The
physical motivation of our proposal is that the moment of inertia
carries richer information about the mass distribution inside a
star, which certainly affects f-mode oscillations. Hence, the
relationship between the moment of inertia and the f-mode is
likely to be more direct and less EOS-dependent. Besides, as
mentioned in Section 2, for compact stars with a fixed EOS, the
moment of inertia bears a one-one correspondence with the
compactness. To this end, we replace the compactness $M/R$ by
another dimensionless factor $\eta \equiv \sqrt{M^3/I}$, which is
proportional to the compactness for stars with uniform densities
and is termed as the effective compactness.

The scaled f-mode frequencies for various realistic NS and QS are
plotted against the effective compactness $\eta$ in
Figure~\ref{fig:f_mode_real_I}, where the f-mode QNM frequencies
of the compact stars constructed with the following EOS are shown:

\begin{figure}
    \begin{center}
        \includegraphics{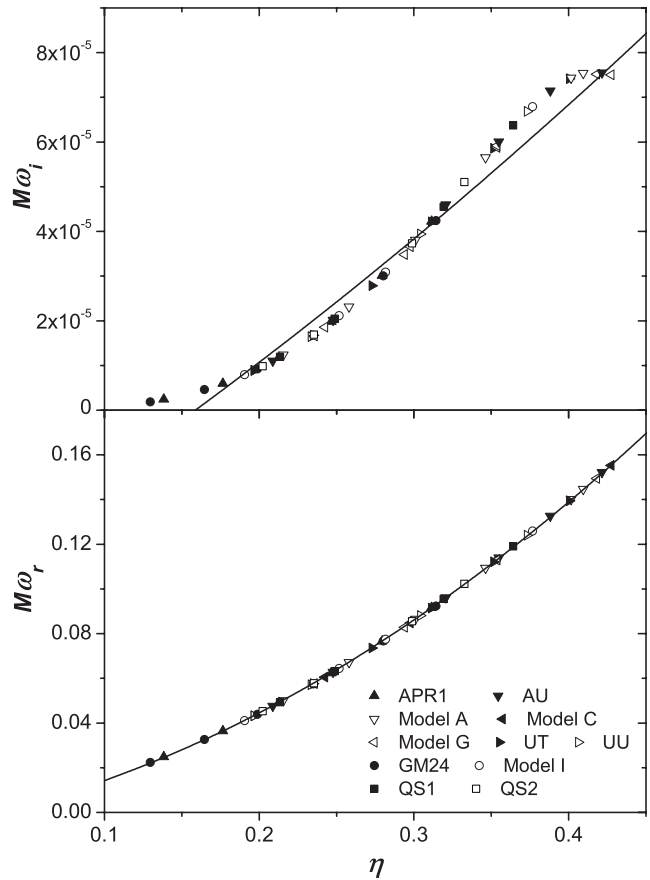}
    \end{center}
    \caption{The scaled frequency of f-mode QNM is plotted against the effective compactness $\eta$.
    The solid lines in the lower and upper panels respectively represent the best quadratic fits
    in $\eta$ to $M \omega_\r$ and $M \omega_\i$. In particular, the solid line in the lower panel is given
    by equation~(\ref{lau_r}).}
         \label{fig:f_mode_real_I}
\end{figure}

\begin{table*}
\caption{\label{uni_quality} Percentage error of different
universal formulae}
\begin{center}
\begin{tabular}{ccccc}
\hline quantity & equation & $\sigma_1$ & $\sigma_2$ & $\sigma_3$ \\
\hline
\hline $\omega_\r$ & (\ref{leung1}) & 5.8 & 4.2  & 21 \\
\hline $\omega_\r$ & (\ref{lau_r}) & 0.37 & 0.56  & 0.70 \\
\hline
\hline $\omega_\i$ & (\ref{leung2}) & 220 & 6.1  & 55 \\
\hline $\omega_\i$ & (\ref{lau_i}) & 1.5 & 2.1  & 1.1 \\
\hline
\end{tabular}
\end{center}
\end{table*}

\begin{enumerate}
\item Nine popular ordinary nuclear matter EOS for NS, including
Model~A \citep{modelA}; Model~C \citep{modelC}; Model~G
\citep{modelG}; Model~I \citep{modelI}; APR~1 \citep{APR}; AU, UU
and UT \citep{AU}; and GM24 \citep[p.~244]{ComStar}, which cover a
wide range of softness (with an adiabatic index $\gamma \sim1.6-4$
in the high-density region). \item Two QS models, QS1 and QS2,
constructed with the MIT bag model where the effects of strange
quark mass and temperature are considered \citep[see,
e.g.,][]{ComStar}. Characteristic physical parameters for QS1
(QS2) are as follows: the bag constant $B$ is given by
$B^{1/4}=154.5~\textrm{MeV}$ ($150.0~\textrm{MeV}$), strange quark
mass $m_s = 150~\textrm{MeV}$ ($100~\textrm{MeV}$), and
temperature $T = 0~\textrm{MeV}$ ($10~\textrm{MeV}$).
\end{enumerate}

\begin{figure}
    \begin{center}
\includegraphics{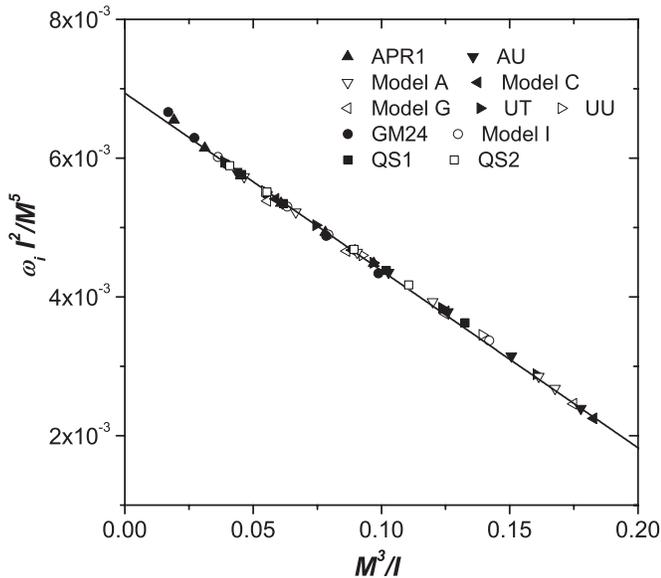}
    \end{center}
    \caption{$\omega_\i I^2/M^5$ is plotted against $\eta^2$ (i.e., $M^3/I$). The solid line is the best linear fit to the distribution (see equation~(\ref{lau_i})). }
    \label{fig:f_mode_imaginary_I}
\end{figure}

It is clearly seen that the data for NS and QS now display almost
identical universal behaviors with high degree of accuracy. In
fact, a quadratic expression in $\eta$ of the form
\begin{equation}\label{lau_r}
M \omega_\r = -0.0047 + 0.133 \eta + 0.0575 \eta^2 \ ,
\end{equation}
suffices to yield a perfect fit to $M\omega_{\r}$ of all stars
(NS and QS). On the other hand, despite the fact that
$M\omega_{\i}$ of all stars fall on a common smooth curve,
quadratic fits in $\eta$ do not work very well in this case.

Motivated by the quadrupole radiation formula of GW and hence the
suggestion that $\omega_\i \propto M^3/R^4$ \citep{Andersson1998},
we plot $\omega_\i I^2/M^5$ against $\eta^2$ (i.e., $M^3/I$) in
Figure~\ref{fig:f_mode_imaginary_I} and see that all data points
now fall on a single straight line with high accuracy:
\begin{equation}\label{lau_i}
I^2 \omega_\i / M^5 = -0.00694 + 0.0256 \eta^2\,.
\end{equation}
Therefore, we conclude that moment of inertia is a good physical
parameter to unveil the universality inherent in the f-modes of
compact stars.

To compare the accuracy of various proposed universal curves for
the f-modes quantitatively, we evaluate the root-mean-square
percentage error $\sigma$ defined by
\begin{equation}\label{}
 \sigma \equiv \sqrt{\frac{1}{N} \sum_i^N \Big(1- \frac{\hat{y}_i}{y_i}
 \Big)^2}~.
\end{equation}
Here $N$ is the total number of f-mode data considered in the
sample. $y_i$ corresponds to the exact value of the physical
quantity $y$ (e.g., $M\omega_{\r}$ and $M\omega_{\i}$) of the
$i$-th f-mode data. $\hat{y}_i$ is the approximated value of $y$
inferred from the universal curves discussed above. In
Table~\ref{uni_quality}  we show the percentage error $\sigma$ in
$\omega_\r$ and $\omega_\i$ for the data obtained from different
universal curves and different samples of f-mode data, namely
those of (i) NS shown in Figure~\ref{fig:f_mode_real_I}
($\sigma_1$); (ii) NS considered in \citet{Andersson1998}
($\sigma_2$); and (iii) QS shown in
Figure~\ref{fig:f_mode_real_I} ($\sigma_3$).

It is clearly shown in Table~\ref{uni_quality} that the universal
curves  (\ref{lau_r}) and (\ref{lau_i}) proposed in the present
paper, which considers mass and moment of inertia as two crucial
parameters characterizing f-mode pulsations of compact stars, work
nicely with high precisions. For $\omega_\r$, the error is less
than 1\%, while for $\omega_\i$, the error is only about 1-2\%. In
particular, such high degree of accuracy prevails for both NS and
QS, whereas the universal curves using conventional compactness
fail to yield satisfactory result for QS. Thus, under the
assumption that the f-mode GW signals from a compact star could be
detected, we can then infer the mass and moment of inertia of the
star from its f-mode using the universal curves (\ref{lau_r}) and
(\ref{lau_i}), without any prior knowledge of its EOS and
composition. We shall establish a feasible inversion scheme to
achieve such purpose in the following section.

Before introducing the inversion scheme, we perform an independent
test here to show that equations (\ref{lau_r}) and (\ref{lau_i})
are robust and accurate for other commonly considered EOS. We
consider the f-mode QNM of NS constructed  with five modern EOS of
nuclear matter, namely N1H1 \citep{Balberg:1997yw}, BBB2
\citep{Baldo:1997ag}, BPAL12 \citep{BPAL12}, FPS \citep{FPS} and
SLy4 \citep{DH2001}, which have  not been used in the derivation
of equations (\ref{lau_r}) and (\ref{lau_i}).  The
root-mean-square percentage errors $\sigma$ for these f-modes  are
just $0.25\%$ and $0.95\%$, respectively. This once again
demonstrates the robustness of our finding reported here.


\begin{table*}
\caption{\label{invert_exp} Percentage error of our inversion
scheme}
\begin{center}
\begin{tabular}{|cc|ccc|}
\hline EOS & $M$ &  $\delta M/M$ & $\delta R/R$ & $\delta I/I$ \\
\hline
\hline AU& 0.8 & -0.056 & -2.6 & -0.031 \\
\hline AU     & 1.0 &  -0.13 & -0.67 & -0.16 \\
\hline AU     & 1.6  & 4.0 & 2.1 & 6.9 \\
\hline APR1 & 0.8 & -0.066 & -5.1 & -0.23 \\
\hline APR1          & 1.2  & -0.21 & -1.6 & -0.19 \\
\hline APR1          & 1.6 & 0.50 & -0.11 & 0.78 \\ \hline
\hline EOS A & 1.535  & 0.40 & -1.2 & 0.27 \\
\hline  EOS A          & 1.328  & 1.1 & -0.46 & 1.8 \\
\hline EOS B & 1.405 &  -0.17 & -3.5 & 0.54 \\
\hline EOS B            & 0.971  & 1.4 & -4.1 & 2.7 \\
\hline GM24 & 1.536  & -2.2 & -4.9 & -2.6 \\
\hline GM24      &1.405       & 2.5 & -1.1 & 5.1 \\ \hline
\hline  QS1    & 0.8  & -0.096 & 18 (0.53)& -0.18 \\
\hline  QS1           & 1.4  & 0.93 & 12 (1.3)& 1.3 \\
\hline  QS2    & 1.0  & -0.17 & 17 (0.72) & -0.3 \\
\hline  QS2           & 1.6  & 1.6 & 11 (1.7) & 2.5 \\
\hline
\end{tabular}
\end{center}
\end{table*}

\section{\label{Application} Inversion scheme}
Assuming that both $\omega_\r$ and $\omega_\i$ of the f-mode of a
compact star have been obtained from GW observations, we can then
solve the simultaneous equations (\ref{lau_r}) and (\ref{lau_i})
to determine the mass $M$ and the moment of inertia $I$ of the
star.

Moreover, making use of equation~(\ref{Lat}) proposed by
\citet{Lattimer:2005p7082}, we can estimate the radius $R$ with
the input of $M$ and $I$. Since all these three equations hold
only approximately, the inferred values of $M$, $R$ and $I$ are
expected to deviate from the exact values by $\delta M$, $\delta
R$, and $\delta I$. As shown in Table~\ref{uni_quality},
percentage errors in $\omega_\r$ and $\omega_\i$ predicted by
equations (\ref{lau_r}) and (\ref{lau_i}) are small and usually
less than a few percent. It is reasonable to expect that the
percentage errors $\delta M/M$ and $\delta I/I$ are also small and
such inversion scheme can lead to accurate determination of the
mass and the moment of inertia of the star.

To validate the scheme outlined above and test its accuracy, we
inferred $M$, $R$ and $I$ numerically from $\omega_\r$ and
$\omega_\i$ for three categories of compact stars, namely NS
considered in Figure~\ref{fig:f_mode_real_I} (with AU and APR1
EOS), NS considered in \citet{Andersson1998} (with A, B and GM24
EOS), and QS (QS1 and QS2). Table~\ref{invert_exp} shows the
percentage errors in mass, radius and moment of inertia of our
scheme for these stars. In general, the inversion scheme works
very well for NS and all three physical quantities can be
determined up to a few percent or better. Our scheme also
reproduces accurate values of $M$ and $I$ for QS. However, the
inferred value of $R$ is less satisfactory ($\sim 10-20$\%). This
is understandable as the empirical relation (\ref{Lat}) only
works for NS, but not QS. If, however, there is other independent
evidence (e.g., via the observed thermal X-ray emission)
suggesting that the star might be a QS, we could then use
equation~(\ref{Haensel_QS}) proposed by \citet{Bejger:2002p8392}
to infer a better value of $R$ from $M$ and $I$. The percentage
errors $\delta R/R$ obtained in this way are enclosed by
parentheses in Table~\ref{invert_exp}. We see that the percentage
error in $R$ is decreased from about 10\% to 1\% level. In
summary, we expect that the inversion scheme proposed in the
present paper is applicable regardless of the nature of the
compact star in consideration.

\section{\label{Conclusion}Conclusion and discussions}
In this paper we study the universality embedded in f-mode
pulsations of compact stars and propose to use it to determine the
physical parameters (including mass, radius and moment of inertia)
of a compact star from which the f-mode gravitational wave signal
is detected. We first establish a pair of empirical equations
(\ref{lau_r}) and (\ref{lau_i}) that can predict the frequency and
the damping rate of f-modes from $M$ and $I$ with good accuracy
for both NS and QS. We then apply such discovery to develop an
inversion scheme to infer the mass, radius and moment of inertia
of a compact star from its f-mode. In particular, the scheme is
shown to work well for both NS and QS.

The universality discovered in the present paper takes moment of
inertia into the consideration, replacing the role of radius used
in previous attempts
\citep{Andersson1996,Andersson1998,Ferrari,Ferrari_prd,Tsui05:1}.
It is worth noting that the radius of a star is sensitive to the
low-density part of its EOS, whereas f-mode frequencies are
expected to be dominated by dynamical behavior in the high-density
regime. It is therefore physically appealing to use moment of
inertia, which measures global mass distribution, to study f-mode
oscillations. As clearly shown in Table~\ref{uni_quality}, the
replacement of $R$ with $I$ indeed leads to a much improved
universal behavior. As it is arguable that the accuracy of the
inversion scheme directly reflects the quality of the universal
behavior upon which the scheme is based, the universality
discovered in the present paper leads to an accurate inversion
scheme.

Finally, it is worthy of mentioning that the first
gravitational-wave search sensitive to the f-modes has recently
been carried out by the LIGO detectors \citep{abbott-2008sgr}.
While the current detectors have not yet detected gravitational
waves directly, it is plausible that the advanced LIGO detectors,
which have a factor of 10 improvement on the sensitivity, might
detect the f-mode gravitational wave signals from compact stars.
The inversion scheme proposed in this work will provide a unifying
way to infer the mass, radius and moment of inertia of both NS and
QS accurately. Of course in reality the situation is often
complicated by the presence of magnetic field and rotation, in the
future we shall consider the impacts of these factors on our
proposal.

\begin{acknowledgments}
This work is supported in part by the Hong Kong Research Grants
Council (Grant No: 401807) and the direct grant (Project ID:
2060330) from the Chinese University of Hong Kong. We thank J.
Wu, Y.~J. Zhang and P.~O. Chan for helpful discussions and
assistance.
\end{acknowledgments}

\pagestyle{plain}
\newcommand{\noopsort}[1]{} \newcommand{\printfirst}[2]{#1}
  \newcommand{\singleletter}[1]{#1} \newcommand{\switchargs}[2]{#2#1}

\newpage

\end{document}